\documentclass[aps,prb,amsmath,amssymb,]{revtex4}
\usepackage{graphicx}
\usepackage{amsmath}
\usepackage{amsfonts}
\usepackage{amssymb}
\usepackage{wasysym}
\usepackage{epsfig}

\newcommand{\beq}{\begin{equation}}
\newcommand{\eeq}{\end{equation}}
\newcommand{\beqa}{\begin{eqnarray}}
\newcommand{\eeqa}{\end{eqnarray}}

\newcommand{\rr}{{\bf r}}

\newcommand{\p}{{\bf p}}

\newcommand{\om}{\omega}

\begin{document}

\title{Vortex structures in  rotating Bose-Einstein condensates}
\author{ S. I. Matveenko$^{1,2}$, D. Kovrizhin$^{3}$, S. Ouvry$^{2}$, and G. V. Shlyapnikov$^{2,4}$}
\affiliation{\mbox{$^1$L.D. Landau Institute for Theoretical Physics, Kosygina Str. 2, 119334, Moscow, Russia}\\
\mbox{$^2$ Laboratoire de Physique Th\'eorique et Mod\'eles Statistiques, Universit\'e Paris
 Sud, CNRS,}\\\mbox{91405~Orsay, France} \\
\mbox{$^3$Theoretical Physics, Oxford University, 1 Keble road, OX1 3NP, Oxford, UK}\\
\mbox{$^4$Van der Waals-Zeeman Institute, University of Amsterdam, Valckenierstraat 65/67,}\\\mbox{1018 XE Amsterdam, The Netherlands}}
\date{\today}

\begin{abstract}
We present an analytical solution for the vortex lattice in a rapidly rotating trapped Bose-Einstein
condensate (BEC) in the lowest Landau level and discuss deviations from the Thomas-Fermi density profile. This solution is exact in the limit of a large number of
vortices and is obtained for the cases of circularly symmetric and narrow channel geometries.
The latter is realized when the trapping frequencies in the plane perpendicular to the rotation axis are different from each other and the rotation frequency is equal to the
smallest of them. This leads to the cancelation of the trapping potential in the direction of the weaker confinement and makes the system infinitely elongated in this direction.
For this case we calculate the phase diagram as a function of the interaction strength and rotation frequency and identify the order of quantum phase transitions between the
states with a different number of vortex rows.
\end{abstract}

\maketitle

\section{Introduction}

Rapidly rotating Bose-condensed gases constitute a novel class of many-body systems where the ground state
 properties are governed by a collective behavior of nucleated vortices \cite{coop,revfet}. A harmonically
 trapped dilute Bose-Einstein condensate (BEC) strongly confined in the $z$ direction, is essentially
 two-dimensional in the $(x,y)$ plane. When the rotation frequency along the $z$ axis becomes close to the
 trapping frequencies in the $x$ and $y$ directions, the BEC gas can be described as a system of interacting
 bosons in the lowest Landau level. The single-particle Hamiltonian is similar to that of a charged particle
 in a strong magnetic field, and the regime of fast rotation of neutral bosons presents an analogy
 with Quantum Hall Effect. Due to the presence of remaining harmonic trapping, the lowest Landau level (LLL)
 is not degenerate. However, analytic properties of the LLL wave functions generate an effective
 long-range interaction between the bosons, which results in an interesting physics.

If the rotation frequency is not very close to the trap frequency, then the number of vortices is much
 smaller than the number of particles. Under these conditions the system is in the so-called
mean-field Quantum Hall regime and can be described by a macroscopic wavefunction $\Psi(\bf r)$ in the lowest
 Landau level. In this limit the vortices generically arrange themselves in a lattice. An increase in the
 rotation frequency increases the number of vortices and eventually it becomes comparable with the number of
 particles. This leads to melting of the vortex lattice and to the appearance of strongly correlated
 states \cite{coop,revfet}. The ``mean-field Quantum Hall regime'' for trapped bosons has been introduced by Ho
 \cite{ho} and studied in a number of papers where the vortex lattice structures have been obtained
 numerically in the case of a circularly symmetric trapping potential \cite{num1,num2,num3,aft2}.

In this paper we consider a rotating BEC in the lowest Landau level in the mean-field regime and obtain an
 analytical solution for the vortex lattice of the harmonically trapped symmetric 2D gas. This solution is
 exact in the limit of a large number of vortices, and we discuss deviations from the Thomas-Fermi density profile.
We then turn to the case of the ``narrow channel''
 geometry, which is realized when the confining frequencies in the $x$ and $y$ directions are different, and
 the rotation frequency is equal to the smallest of them. Then, in the rotating frame, the gas becomes
 extremely  elongated in the direction of the smaller frequency, as has been demonstrated in the ENS
 experiment with thermal bosons \cite{rosenbusch}. This is an extreme case of a rapidly rotating 2D gas in an
 asymmetric harmonic potential, discussed in relation to the density profile of the gas and the density of vortices
in Ref.~\cite{fet}. Some vortex structures of the asymmetric rapidly rotating BEC have been discussed and calculated in
 Refs.~\cite{fet0,oktel,aft1,aft3}. In the narrow channel geometry, the excitation spectrum of a weakly interacting
 BEC without vortices exhibits a ``roton-maxon''  structure \cite{gora}. The phase transition to the state
 with a vortex row occurs when the roton energy reaches zero under an increase in the rotation frequency or
 in the strength of interaction between the bosons. A further increase of these quantities increases  the
 number of vortex rows through a set of quantum phase transitions \cite{gora,sanchez}. We classify these
 transitions and find  an analytical solution for the vortex lattice in the narrow channel, which is exact in
 the limit of a large number of vortex rows.

\section{Gross-Pitaevskii equation in the lowest Landau level. Solution for a symmetric harmonic potential}

Consider a system of bosonic neutral atoms strongly confined in the $z$ direction by an external trapping potential with  frequency $\omega_z$ such that the bosons are in the
ground state of the $\omega_z$ harmonic well and  become essentially two-dimensional in the $(x,y)$-plane. The bosons are confined in this plane by a harmonic trapping potential
$V({\bf r})$, with ${\bf r}=\{x,y\}$, and the trap is rotating around the $z$ axis with frequency $\Omega$. In the mean-field Quantum Hall limit, we assume to zero order that all
particles are in the same macroscopic quantum state described by the wavefunction $\psi({\bf r})$. In the rotating frame the Gross-Pitaevskii equation for $\psi({\bf r})$ reads:
\beq
\frac{\hat{\p}^2}{2 m} \psi + g |\psi|^2 \psi  +V(\rr) \psi  - \Omega \hat{L}_z \psi = \mu \psi,
\label{gp}
\eeq
where $\hat{\p}$ is the momentum operator, $m$ is the particle mass, $\hat{L}_z$ is the operator of the orbital angular momentum, $\mu$ is the chemical potential, and $\psi$ is
normalized to the total number of particles $N$. Equation (\ref{gp}) is obtained for a short-range interaction between particles, and the 2D coupling constant $g$ can be
expressed through the 3D scattering length $a_s$. If the harmonic oscillator length in the $z$ direction, $l_z=\sqrt{\hbar/m\omega_z}$, is much larger than $|a_s|$ and the
characteristic radius of interparticle interaction, then we have \cite{gora1}:
\beq
g=\frac{2\sqrt{2\pi}\hbar^2a_s}{ml_z}.
\label{g}
\eeq

We will study Eq.~(\ref{gp}) projected onto the lowest Landau level. A general procedure of obtaining the projected equation is described in the Appendix, and here we outline the
method.

The single-particle Hamiltonian for rotating neutral atoms is equivalent to the Hamiltonian of a charged particle in a uniform magnetic field $B$ along the $z$ axis. The field is
such that half the cyclotron frequency $\omega_c=B/2m$ (in units of charge divided by the light velocity) is identified with the rotation frequency $\Omega$, and the
vector-potential in the symmetric gauge is ${\bf{A}}=[{\bf B}\times\rr]/2=m[{\bf{\Omega}}\times\rr]$. In the case of a symmetric external harmonic potential $V(r)=m\omega^2r^2/2$
the single-particle Hamiltonian reads:
\beq
H=\frac{\hat{\p}^2}{2 m} - \Omega \hat{L}_z  + \frac{1}{2}m \om^2 r^2 =
\frac{1}{2 m}{(\hat{\p} - e {\bf{A}} /c)^2} + \frac{1}{2}m (\om^2 -\Omega^2)  r^2,
\label{Hsingle}
\eeq
At the critical rotation frequency $\Omega=\om$ the residual confining potential vanishes, and the harmonic oscillator length $l=(\hbar/m\omega)^{1/2}$ of the initial trapping
potential $V(r)$  coincides with the ''magnetic length'' $(\hbar/m\Omega)^{1/2}$. One then has an ''infinite plane'' geometry actively studied with respect to the ground state of
interacting bosons \cite{sw,cwg,paredes,regnault,mash}.

Below the critical rotation frequency, $\Omega<\om$ and $(\omega-\Omega)\ll\Omega$, the energy eigenstates are associated with the Landau levels of a charged particle in the
uniform magnetic field, and the
presence of the residual confining potential lifts the LLL degeneracy. A complete set of LLL eigenfunctions is given by
\beq
\Psi_n(z,{\bar z})=\frac{z^n}{l\sqrt{\pi n!}}\exp\left[-\frac{z{\bar z}}{2}\right],
\label{Psin}
\eeq
with $z,{\bar z}=(x\pm iy)/l$, and $n$ being a non-negative integer. An arbitrary function in the LLL can be written as a linear superposition of the LLL eigenstates and
represented in the form:
\beq
\Psi(z,{\bar z}) = \frac{f(z)}{l}\exp\left[-\frac{z{\bar z}}{2}\right],
\label{Psizz}
\eeq
where $f(z)$ is an analytic function of $z$. The projection operator onto the LLL is written as
\beq
\hat P=\sum_n   |n \rangle  \langle n|,
\label{hatP}
\eeq
where $\langle z|n\rangle = \Psi_n(z,{\bar z})$  is given by Eq.~(\ref{Psin}).
Acting with the operator $\hat P$ on an arbitrary  function $\phi(z,{\bar z})=[f(z,\bar{z})/l]\exp(-z{\bar z}/2)$ one obtains:
\beq
\hat P\phi(z,{\bar z})=\sum_{n=0}^{\infty}\int\Psi_n(z,{\bar z})\Psi_n^*(z^{\prime},{\bar z}^{\prime})\phi(z^{\prime},{\bar z}^{\prime})dz^{\prime}d{\bar z}^{\prime}=[\tilde
f(z)/l]\exp(-z{\bar z}/2),
\label{acting}
\eeq
with $dz'd\bar{z'}=dx'\,dy'/l^2$, and an analytic function $\tilde{f}(z)$ which is given by
\beq
\tilde{f}(z) =\frac{1}{\pi}\int dz'd\bar{z'} f(z',\bar{z'})\exp(z\bar{z}'-z'\bar{z'}).
\label{tildef}
\eeq
This formalism was introduced by Bargmann \cite{bargmann} and developed by Girvin and Jach \cite{girvin} in relation to Quantum Hall physics.

In the case of interacting bosons one can still project the many-body Hamiltonian onto the lowest Landau level provided that the interactions are much smaller than the cyclotron
gap,
i.e. $n_{2D}g\ll2\hbar\Omega$, where $n_{2D}$ is the two-dimensional particle density \cite{coop,revfet}. Acting with the LLL projector $\hat P$ onto the Gross-Pitaevskii equation
(\ref{gp}) results in the projected equation (see Refs. \cite{num2,aft1} and Appendix):
\beq
\hbar(\omega-\Omega)z\partial_z f(z)+\frac{Ng}{\pi l^2}\int dz'd\bar{z'}|f(z')|^2 f(z')\exp(z\bar{z}'-2 z'\bar{z}')=\tilde\mu f(z),
\label{pgp}
\eeq
where $\tilde\mu=\mu-\hbar\omega$, and the function $[f(z)/l]\exp(-z{\bar z}/2)$ is normalized to unity. Equation (\ref{pgp}) has a simple solution
\beq
f_n(z)=\frac{z^n}{\sqrt{\pi n!}},
\eeq
which corresponds to the chemical potential and energy per particle given by
\begin{eqnarray}
&&\mu_n=\hbar(\omega-\Omega)n+\frac{Ng}{2\pi l^2}\frac{(2n)!}{(n!)^2 2^{2n}},  \label{mun} \\
&&\frac{E_n}{N} =\hbar(\omega-\Omega)n+\frac{Ng}{4\pi l^2}\frac{(2n)!}{(n!)^2 2^{2n}}.  \label{energy}
\end{eqnarray}
For $n = 0$, equation (\ref{energy}) describes  the ground state without vortices, and for $n\ne0$ it gives excited states with a (multicharged) vortex at the origin.
The LLL approximation is valid when $E_n\ll\hbar\omega N$. The spectrum $E_n$ (\ref{energy}) has a roton shape with a local minimum at a certain value of $n$. In the limit of large
$n$ we have
$n! \approx \sqrt{2 \pi n}(n/e)^n$, and Eq.~(\ref{energy}) is reduced to
\beq
\frac{E_n}{N}=\hbar(\omega-\Omega)n+\frac{Ng}{4\pi l^2}\frac{1}{\sqrt{\pi n}}.
\label{energylargen}
\eeq
The local energy minimum is obtained for
\beq
n=n_0=\frac{1}{4\pi}\left(\frac{Ng}{\hbar(\omega-\Omega)}\right)^{2/3},
\label{n0}
\eeq
and from Eq.~(\ref{energylargen}) we find
\beq
E_{n_0}=\frac{3}{4\pi}\left(\frac{Ng}{l^2}\right)^{2/3}[\hbar(\omega-\Omega)]^{1/3}.
\label{energyn0}
\eeq
The giant vortex state at $n=n_0$ is a metastable state and it can have a relatively long lifetime. One can think of creating this state in dynamical studies and identifying it
through the measurement of the density profile \cite{foot}.

For $Ng/l^2\gg\hbar(\omega-\Omega)$ the ground state represents a vortex lattice. At the critical rotation frequency $\Omega = \omega$, one finds an exact solution describing this
lattice. Consider the function
\begin{equation}
f_0(z) = \frac{(2v)^{1/4}}{\sqrt{S}}\vartheta_1 \left(\pi z /b_1, \tau \right)\exp(\pi z^2/2v_c),
\label{f}
\end{equation}
where $S$ is the surface area, $\tau=u+iv$, and $\vartheta_1$ is the Jacobi theta-function \cite{abramowitz} given by
\begin{equation}
\label{theta}
\vartheta_1 ( \zeta, \tau )=\frac{1}{i}\sum_{n= -\infty}^{\infty}(-1)^{n}\exp\{i\pi\tau(n + 1/2)^2+2i\zeta (n+1/2)\}.
\end{equation}
The  Jacobi theta-functions are analytic in the  complex plane and have zeros at the points $\zeta =n \pi + m \pi \tau$,
where $n,m $ are integers. The function $f_0 (z)$ vanishes at the lattice sites $n b_1 + m b_2$, with $b_2 = b_1 \tau$. These points correspond to the vortex
locations.
The parameter $b_1$ can always be chosen real so that  the area  of the unit cell is $v_c = b_1^2 v$. The absolute value of $f_0(z)$ is polar symmetric
\beq
|f_0(z)|\sim \exp \left[\frac{\pi(x^2 + y^2)}{2 v_c l^2}\right].
\label{polarsymmetric}
\eeq
For any lattice with a fixed elementary cell area one has ${v_c} = \pi$, and the normalization coefficient in Eq.~(\ref{f}) is chosen such that the function $[f_0(z)/l]\exp(-z{\bar
z}/2)$ is normalized to unity. The function $f_0(z)$ is an exact solution of Eq.~(\ref{pgp}) for $\omega=\Omega$. It has  a constant envelope and
a periodic vortex structure. The minimum energy is obtained for the triangular lattice, where $\tau = \exp 2\pi i/3$, $v=\sqrt{3}/2$, and $b_1^2 = 2 \pi/\sqrt{3}$. The chemical
potential is then
given by
\[
\mu=\frac{\alpha Ng}{l^2},
\]
with $\alpha=(3^{1/4}/2)\sum_{b,c}(-1)^{mp}\exp\{-\pi^2(b^2+c^2)/4b_1^2\}=1.1596$, and $b=2m$, $c=2p$ being even integers.

In the case of $\omega>\Omega$, a general solution of equation (\ref{pgp}) can be represented as
\beq
f(z) = (2v)^{1/4}\sum_{n= -\infty}^{\infty} (-1)^n  \hat{g}(a) \tilde{q}^{a^2}   \exp\left[\frac{i \pi}{b_1}az+\frac{z^2}{2}\right],
\label{symf}
\eeq
where  $a=2n+1$ are odd integers, $\tilde{q} = \exp[i \pi \tau /4]$, and $\hat{g}(a)$ is a  differential operator acting on  $a$. For $\hat{g}(a)\equiv 1$ one recovers the solution
(\ref{f}). Substituting the trial function (\ref{symf}) into equation (\ref{pgp}) for a triangular-like lattice (the lattice that becomes exactly triangular for $\omega=\Omega$) we
obtain:
\begin{eqnarray}
&&\left\{(\mu^* - \hat{A}^+ \hat{A})\hat{g}(a) - \frac{3^{1/4}\beta}{\sqrt{2}}\sum_{b, c} \hat{g}(a-b)\hat{g}(a-c)\overline{\hat{g}(a - b - c)} \exp\left[-\frac{\pi^2}{4 b_1^2}(b^2
+ c^2)\right](-1)^{m
p}
\right\}    \nonumber \\
&&\times\tilde{q}^{a^2}\exp\left[\frac{i \pi}{b_1}az\right]=0.
\label{mainsym}
\end{eqnarray}
Here $\beta=Ng/(l^2\hbar(\omega-\Omega))$, $\mu^*=\tilde\mu/\hbar(\omega-\Omega)$, and we introduced the operators
 \[
 \hat{A} , \hat{A}^+ =  \frac{ \pi}{2 b_1} a    \pm \frac{b_1}{\pi} \frac{\partial}{\partial a}.
 \]
 For large $\beta$ we have $\mu^*\gg 1$ and an approximate solution for $\hat g(a)$, which describes the vortex structure with a high accuracy, turns out to be
\beq\label{energybis}
\hat{g}(a) = \frac{1}{\sqrt{\alpha\beta}}  \sqrt{R^2 - \hat{A}^+ \hat{A}}\,\,\Theta(R^2-\hat A^{\dagger}\hat A),
\eeq
where $\Theta$ is the Heaviside theta function, $R=\sqrt{\mu^*}$, and we will see below that it is the radius of the condensate cloud in units of $l$.
Substituting the solution (\ref{energybis}) into Eq.(\ref{symf}) we obtain after some algebra:
\beq
f(z) = \frac{(2v)^{1/4}}{\sqrt{\alpha\beta}} \sum_{n = -\infty}^{\infty} \sum_{k = 0}^{[R^2]} (-1)^{[n(n-1)/2]}\,\sqrt{R^2 - k}\,\,\frac{ (i z)^k}{2^{k/2} k!} H_k\left
(\sqrt{\frac{\pi v}
{2}} (2n+1)\right)\exp\{-\pi v (2 n + 1)^2/4\},
\label{ffin}
\eeq
where $H_k(w)$ are Hermite polynomials.

Equation (\ref{energybis}) is obtained taking into account that the leading contribution to the sum over $b$ and $c$ in Eq.~(\ref{mainsym}) comes from small values of $m$ and $p$,
since already the contributions of terms with $|m|\geq 2$ or $|p|\geq 2$ are exponentially small. Provided that the dependence  $\hat{g}(a)$ is smooth, which is the case for large
$R$,  we may consider large $a$ and omit $b$ and $c$ in the arguments of the $\hat g$ operators in Eq.~(\ref{mainsym}).  This immediately gives Eq.~(\ref{energybis}).
From the condition that the function $[f(z)/l]\exp(-z{\bar z})$ is normalized to unity we find a relation
$R=(2\alpha\beta/\pi)^{1/4}$, in agreement with Refs. \cite{num2,aft1}.

\begin{figure}[htb]
 \includegraphics[width=4.0in]{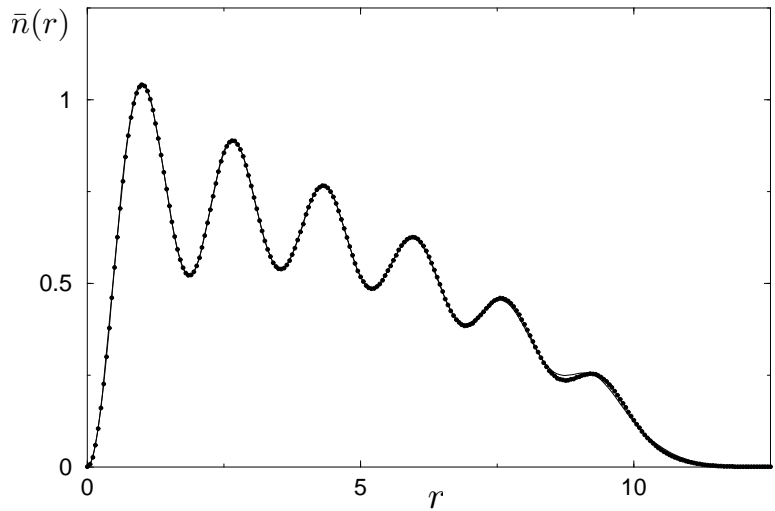}
 \caption{Angular-averaged  density $\bar{n}(r)$  in units of $n_{2D}$ versus $r$  (in units of $l$) for $ R=11$. The solid curve shows the result of Eq.~\ref{n-r}, and filled
circles the results of the variational calculation (see text). }
\label{sym_r11}
 \end{figure}
 \begin{figure}[htb]
\centering
\includegraphics[width=4.0in]{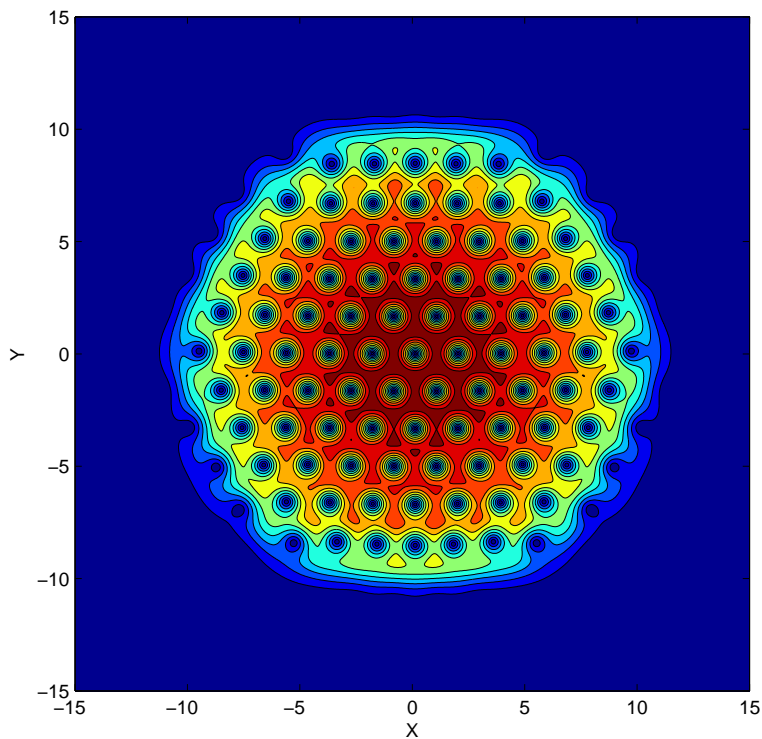}
\caption{Condensate wave-function $|\psi(x,y)|^2$ for $ R=11$. Coordinates $x$ (horizontal line) and $y$ (vertical line) are given in units of $l$.}
\label{mu120}
\end{figure}
\begin{figure}[htb]
 \includegraphics[width=4.0in]{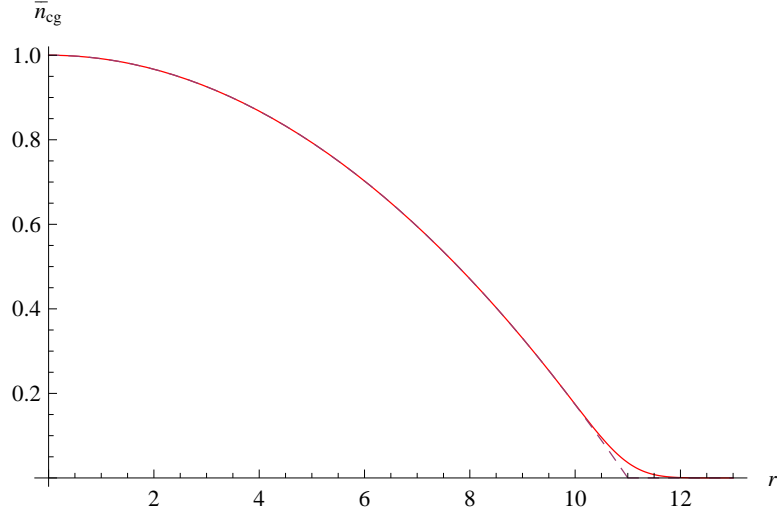}
 \caption{The coarse grained density in units of ${\bar n}_{2D}$ versus $r$, calculated from Eq.~(\ref{cgsym}) for $R=11$. The dashed curve shows the Thomas-Fermi
inverted-parabola
shape (\ref{ncTF}), and $r$ is given in units of $l$.}
 \label{fig00}
 \end{figure}

For the angular-averaged particle density, i.e. the density averaged over the azimuthal angle $\varphi$ ($z = r \exp i\varphi$) we then have:
\beqa
&&\bar{n}(r) = N\int |\psi(z,{\bar z})|^2 \frac{d \varphi}{2\pi} ={\bar n}_{2D}(2v)^{1/2}\sum_{k = 0}^{[R^2]}\,\, \sum_{n,m = -\infty}^{\infty}
\left(1 - \frac{k}{R^2}\right)\, (-1)^{\{[n(n-1)+m(m-1)]/2]\}}\,\frac{ r^{2k}}{2^{k} (k!)^2} \nonumber \\
&& H_k \left(\sqrt{\frac{\pi v}{2}} (2n+1)\right) H_k\left (\sqrt{\frac{\pi v}{2}} (2m+1)\right)\exp\{-\pi v (2 n + 1)^2/4-\pi v (2 m + 1)^2/4\}\exp(-r^2),
\label{n-r}
\eeqa
where ${\bar n}_{2D}=2N/\pi l^2R^2$ is a characteristic 2D density in the central part of the cloud. The angular-averaged density calculated from Eq.~(\ref{n-r})  for $R=11$ is
shown in Fig.~\ref{sym_r11}.  In the entire region of $r$ it coincides with the numerical result obtained by  expanding the condensate wavefunction in terms of the
single-particle LLL states (\ref{Psin}) and using a variational approach for finding the coefficients of the expansion. This demonstrates a very high accuracy of our analytic
solution.  The structure of the vortex lattice for $R=11$ is shown in Fig.~\ref{mu120}.

The angular-averaged density represents oscillations on a length scale of the magnetic length $l$, on top of a slowly varying envelope. Averaging the density over the oscillations,
that is averaging ${\bar n}(r)$ (or just the density $n({\bf r})$) over a distance scale much larger than $l$, gives the coarse grained density:
\begin{equation}             \label{ncggen}
{\bar n}_{cg}(x,y)=\frac{1}{{\cal L}^2}\int_{0}^{{\cal L}}d\delta x\int_0^{{\cal L}}d\delta y\,n(x+\delta x,y+\delta y)\propto |f(z+\delta z)|^2\exp[-(z+\delta z)({\bar
z}+\delta{\bar
z})],
\end{equation}
where $R\gg {\cal L}\gg 1$. Using the function $f(z)$ of Eq.(\ref{symf}) we have:
\begin{eqnarray}
&&n_{cg}(x,y)\propto \frac{1}{{\cal L}^2}\int_{0}^{{\cal L}}d\delta x\int_0^{{\cal L}}d\delta y\sum_{n,m=-\infty}^{\infty}(-1)^{n+m}\hat g(a_{n})\overline{\hat g(a_{m})} \nonumber
\\
&&\exp\left\{\frac{i\pi\tau a_n^2}{4}-\frac{i\pi\tau^*a_m^2}{4}+\frac{i\pi}{b_1}(x+\delta x)(a_n-a_m)-\frac{\pi}{b_1}(y+\delta y)(a_n+a_m)-2(y+\delta y)^2\right\}, \label{ncggen1}
\end{eqnarray}
with $a_n=(2n+1)$, and $a_m=(2m+1)$. Integration over $d\delta x$ gives $n=m$ and transforms the $y$-dependent part of Eq.~(\ref{ncggen1}) to
$$\frac{1}{{\cal L}}\int d\delta y\sum_{n=-\infty}^{\infty}\exp\left\{-\left[\sqrt{2}(y+\delta y)+\frac{\pi (2n+1)}{\sqrt{2} b_1}\right]^2\right\}.$$
One then clearly sees that the integration over $d\delta y/{\cal L}$ is equivalent to replacing the summation over $n$ by integration. Thus, in order to obtain the coarse grained
density from Eq.~(\ref{n-r}) we have to put $n=m$ and integrate over $dn$. This yields:
\beq
\bar{n}_{cg}(r)={\bar n}_{2D}\sum_{k=0}^{[R^2]}\left (1 - \frac{k}{R^2}\right)\,   \frac{ r^{2k}}{ k!}\exp(-r^2)={\bar n}_{2D}\left(1-\frac{r^2}{R^2}\right)\frac{\Gamma(R^2, r^2)}
{\Gamma(R^2)}+\frac{r^{2R^2}\exp(-r^2)}{R^2 \Gamma(R^2)},
 \label{cgsym}
 \eeq
where $\Gamma (R^2)$ and $\Gamma(R^2,r^2)$ are the Gamma function and  Incomplete Gamma function, respectively.

For $r<R$ and $(R-r)\gg 1$ we may put $\Gamma(R^2,r^2)\simeq\Gamma(R^2)$ in the right hand side of Eq.~(\ref{cgsym}) and omit the second term which gives a correction of the order
of $1/R$ or smaller. This leads to the expected  Tomas-Fermi density profile:
\begin{equation}       \label{ncTF}
{\bar n}_{cg}(r)={\bar n}_{2D}\left(1-\frac{r^2}{R^2 }\right)\,\Theta(R-r).
\end{equation}
For $r>R$ and $(r-R)\gg 1$ we have at large $R$:
$$\Gamma(R^2,r^2)=\left(\frac{1}{r^2-R^2}-\frac{r^2}{(r^2-R^2)^3}\right)(r^2)^{R^2}\exp(-r^2).$$
Then, using an asymptotic expression $\Gamma(R^2)=\sqrt{2\pi/ R^2}\,(R^2)^{R^2}\exp(-R^2)$, we
 obtain that the density decays exponentially:
\begin{equation}      \label{ncexp}
{\bar n}_{cg}(r)=\frac{{\bar n}_{2D}}{\sqrt{2\pi R^2}}\frac{\exp[-2(r-R)^2]}{4(r-R)^2};\,\,{\rm at}\,\,R^{1/3}\gg (r-R)\gg 1,
\end{equation}
and is practically zero for $r\gtrsim R^{1/3}$. The coarse grained density versus $r$ at $R=11$ is displayed in Fig.~\ref{fig00}.
At the Thomas-Fermi border, $r=R$, we have ${\bar n}_{cg}={\bar n}_{2D}/\sqrt{2\pi R^2}$. The validity of the Thomas-Fermi inverted-parabola shape, in general, requires the
inequality $(R-r)\gg 1$. Nevertheless, in Fig.~\ref{fig00} we see that for $R=11$ the Thomas-Fermi formula works well already for $r\leq 10$.

Deviations from the Thomas-Fermi density profile of ${\bar n}_{cg}(r)$ have been studied in Ref. \cite{num2} by using the variational procedure. Here we present an analytic
solution and show that it describes very well the density profile, including the non-Thomas-Fermi part.

\section{Narrow channel geometry}

Let us now consider an anisotropic confining potential $V(\rr) =m(\om_x^2 x^2+\om_y^2 y^2)/2$, with $\om_y<\om_x$. At the critical rotation frequency $\Omega=\om_y$, the
centrifugal force cancels the confining potential in the $y$-direction. One then has a
 quasi-one-dimensional system in the rotating frame, which is usually refered to as the
narrow channel geometry. The system is infinitely elongated in the $y$ direction and is confined by a
 residual transverse potential $m(\om_x^2 - \Omega^2)x^2/2$ in the $x$
direction \cite{gora}.

After the transformation to the Landau gauge, $\Psi\rightarrow\Psi\exp(im\Omega xy/\hbar)$, a
 complete set of eigenfunctions of non-interacting particles in the lowest Landau level
of the narrow channel is given by
\beq
\Psi_n=\left(\frac{2}{\pi}\right)^{1/4}\frac{1}{L^{1/2}\tilde l}\exp\left(ik_n\zeta\frac{\Omega}{\tilde\Omega}\right)\exp\left(-\tilde{y}^2-\frac{k_n^2\Omega^2}
{4\tilde\Omega^2}\right),
\label{setch}
\eeq
where $L$ is the length of the system in the $y$ direction in units of
 $\tilde l=(\hbar/m\tilde\Omega)^{1/2}$, $\tilde\Omega^2=(\omega_x^2+3\Omega^2)/4$, $k_n=2\pi n/L$ with $n$
being an integer, and we introduced dimensionless coordinates
\begin{equation}
\tilde x=-\frac{\tilde\Omega y}{\Omega \tilde l};\,\,\,\,\,\,\tilde y=\frac{x}{\tilde l};\,\,\,\,\,\,\zeta=\tilde x+i\tilde y.  \label{nch}
\end{equation}
Thus, the wavefunction of any state in the LLL can be written in the form:
\beq
\Psi=[f(\zeta)/\tilde l]\exp \left(- \tilde{y}^2\right).
\label{psiLLLch}
\eeq

The projection operator onto the LLL is given by Eq.~(\ref{hatP}), and acting with this operator on an arbitrary function $\phi(\zeta,{\bar \zeta})=[f(\zeta,{\bar \zeta})/\tilde
l]\exp(-\tilde y^2)$ we obtain an analog of equations (\ref{acting}) and (\ref{tildef}):
\begin{eqnarray}
\!\!\!\!\!\!&&\hat{P}\phi(\zeta,{\bar \zeta})=\sum_n\frac{\Omega}{\tilde\Omega}\int\psi_n(\zeta,{\bar \zeta})\psi_n^*(\zeta',{\bar \zeta}')\phi(\zeta',{\bar \zeta}')dx'dy'=[\tilde
f(\zeta)/l]
\exp(-\tilde y^2); \label{actingch}  \\
\!\!\!\!\!\!&&\tilde f(\zeta)=\frac{1}{\pi}\int d\zeta'd{\bar \zeta}'f(\zeta',{\bar \zeta}')\exp\{\zeta \bar{\zeta}'+\zeta'^2 /2-\zeta'\bar{\zeta}'-\zeta^2/2\},  \label{tildefch}
\end{eqnarray}
where $d\zeta'd{\bar \zeta}'=(\tilde\Omega/\Omega)dx'dy'/\tilde l^2$.

\begin{figure}[htb]
 \includegraphics[width=4.0in]{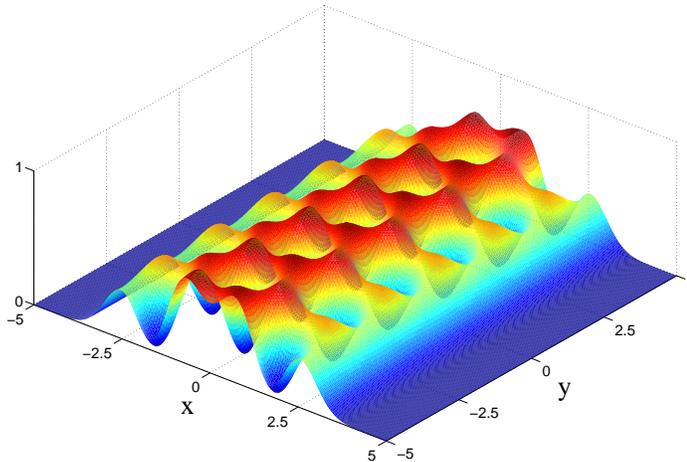}
 \caption{Density profile $|\psi (x, y)|^2$ for $\tilde\beta\simeq 50$. Coordinates $x$ and $y$ are given in units of $\tilde l$, and $|\psi|^2$ in arbitrary units.}
 \label{Fig2}
 \end{figure}

The Gross-Pitavevskii equation projected onto the lowest Landau level in the narrow channel has the form (see Appendix for details):
\beq
-\hbar\omega_0 f''(\zeta) + \frac{Ng}{\pi \tilde l^2}\int d\zeta'd{\bar \zeta}'|f(\zeta')|^2
f(\zeta')\exp\left(-2\zeta'\bar{\zeta}'+\zeta\bar{\zeta}'+\zeta'^2+\frac{\bar{\zeta}'^2}{2}-\frac{\zeta^2}{2}\right)=\tilde\mu f(\zeta),
\label{1d}
\eeq
with $\omega_0=\tilde\Omega(\tilde\Omega^2-\Omega^2)/(2\Omega^2)\ll \Omega$ being proportional to the frequency of the remaining confinement in the $x$ direction,
$\tilde\mu=\mu-\hbar\omega$, and the condensate wavefunction $[f(\zeta)/\tilde l]\exp(-\tilde y^2)$ being normalized to unity.
In analogy with Eq.~(\ref{symf}) let us again  search for the solution of the form
\beq
f(\zeta) =
\frac{(2v)^{1/4}}{\sqrt{L}}\sum_{n=-\infty}^{\infty}  (-1)^n g(2 n +1) \exp\left(i\pi\tau\frac{(2n+1)^2}{4}+\frac{i\pi\zeta(2 n+ 1)}{b_1}\right) ,
\label{subst}
\eeq
where $b_1^2=2\pi/\sqrt{3}$, $v=\sqrt{3}/2$ and $\tau=\exp(2\pi i/3)$.
Equation (\ref{subst}) describes the structure with odd number of vortex rows, with the central row at $y=0$. Using $\vartheta_4$ instead of $\vartheta_1$ in Eq.~(\ref{subst}),
which corresponds to the replacement $(2n+1)\rightarrow 2n$, we obtain structures with an even number of vortex rows.
Substituting Eq.~(\ref{subst}) into Eq.~(\ref{1d}) yields
\beq
\!g(a)\left(\mu^*-\frac{\pi^2 a^2}{b_1^2}\right)=3^{1/4}\sqrt{2\pi}\tilde\beta\sum_{b, c} g(a -b) g(a - c) \overline{g(a - b - c)} \exp\left[-\frac{\pi^2}{4 b_1^2}(b^2 +
c^2)\right](-1)^{mp}
\label{g(a)},
\eeq
where $\tilde\beta = Ng/2\sqrt{\pi}\hbar\omega_0L\tilde l^2$, $\mu^* =\tilde\mu/\hbar\omega_0$, $a=2n+1$, $b=2m$ and $c=2p$ are odd and even integers, respectively. As well as in
the symmetric case, at large $\tilde\beta$ we find an approximate solution for $g(a)$, which describes the vortex structure with a high accuracy:
\beq
g(a) =\left(\frac{1}{2\alpha\sqrt{\pi}\tilde\beta}\right)^{1/2}\frac{\pi}{b_1}\sqrt{4\tilde R^2/\pi v - a^2}\,\, \Theta\left[\frac{2\tilde R}{\sqrt{\pi v}}-a\right],
\label{xyz}
\eeq
where we put $\mu^*=4\tilde R^2$, and it will be seen below that $\tilde R$ is the half-size of the cloud in the $x$ direction (in units of $\tilde l$). From the condition that the
condensate wavefunction is normalized to unity we obtain:
\begin{equation}
\tilde R\approx (3\alpha\sqrt{\pi}\tilde\beta/8)^{1/3}.
\label{tildeR}
\end{equation}
Equation (\ref{xyz}) is obtained by putting the arguments of the $g$ functions equal to $a$ in the sum over $b,c$ in Eq.~(\ref{g(a)}). Similarly to the symmetric case, the
contribution
of terms with high $b$ and $c$ in this sum ($|m|\geq 2$ or $|p|\geq 2$) is very small, except for $n$ very close to the border value $b_1\tilde R/\pi$.
The relative contribution of such $n$ to the sum in
Eq.~(\ref{subst}) decreases with increasing $\tilde R$. Thus, the solution  $f(\zeta)$ (\ref{subst}) with $g(2n+1)$ of equation (\ref{xyz}) becomes exact in the limit of large
$\beta$. The structure of the vortex lattice for $\beta=50$ is displayed in Fig.~\ref{Fig2}. For very large $\tilde\beta$ the number of rows for a triangular-like lattice is
approximately equal to $2\tilde R/(\sqrt{3}b_1/2)\propto
\tilde\beta^{1/3}$ (see next Section). Then, our results lead to the Thomas-Fermi density profile in the $x$ direction for the major part of the cloud, as explained below.
\begin{figure}[htb]
 \includegraphics[width=4.0in]{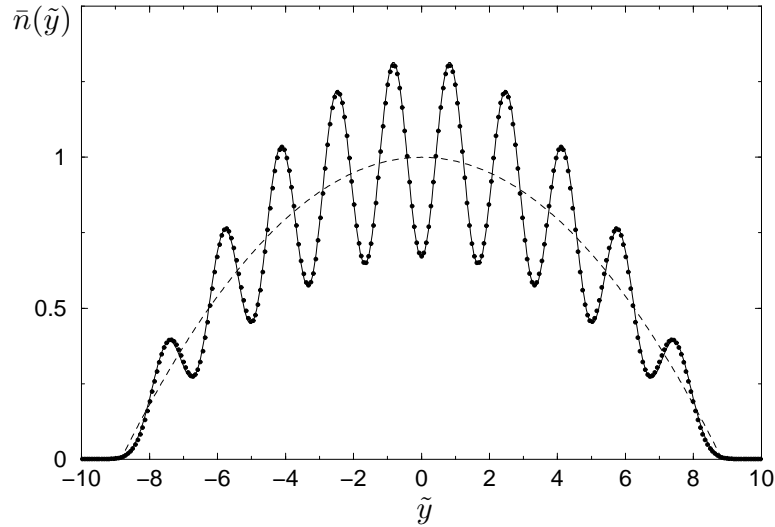}
 \caption{Line-averaged  density $\bar{n}(y)$  in units of $n_{2D}$ versus $\tilde y=x/\tilde l$  for a condensate in the narrow channel for $\beta=900$, $\tilde R=8.8397$. The
solid curve shows the result of Eq.~(\ref{ncg1}), the filled circles indicate the results of the variational calculation (see text), and the dashed curve the Thomas-Fermi
inverted-parabola density profile.}
\label{fig_777}
 \end{figure}
Using equations (\ref{subst}) and (\ref{xyz}) we define the line-averaged density ${\bar n}(\tilde y)$, i.e. the density averaged over the direction $y$ of vortex lines:
\begin{eqnarray}
&&\bar{n}(\tilde y)=\frac{N}{L}\int |\psi(\tilde x,\tilde y)|^2d\tilde x=\frac{3^{1/4}\pi^{3/2}}{2\alpha\tilde\beta b_1^2}\sum_{-n_{max}}^{n_{max}}\left(\frac{4\tilde R^2}
{\pi v}-(2n+1)^2\right)  \nonumber  \\
&&\times\exp\left\{-2 [y +\frac{\sqrt{\pi v}}{2}(2n+1)]^2 \right\},
\label{ncg1}
\end{eqnarray}
with $(2n_{max}+1)\simeq 2b_1\tilde R/\pi$. In Fig.~\ref{fig_777} we compare the result of Eq.~(\ref{ncg1}) with the result obtained by expanding the condensate wavefunction in
terms of the single-particle LLL states (\ref{setch}) and using a variational approach for finding the coefficients of the expansion. The comparison shows a very high accuracy of
the found analytic solution.

\begin{figure}[htb]
 \includegraphics[width=4.0in]{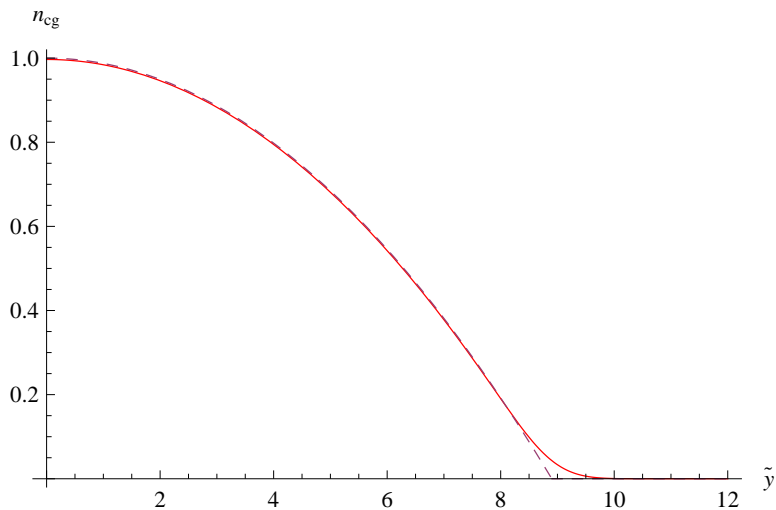}
 \caption{The coarse grained density in units of $n_{2D}$ versus $\tilde y=x/\tilde l$, calculated from Eq.~(\ref{ncg2}) for $\tilde R=8.9$.
 The dashed curve shows the Thomas-Fermi shape.}
 \label{fig0}
 \end{figure}

The line-averaged density ${\bar n}(\tilde y)$ shows oscillations on a length scale $\sim \tilde l$, on top of a slowly varying envelope. Averaging the density over the
oscillations, i.e. averaging ${\bar n}(\tilde y)$ over a distance scale much larger than $\tilde l$, gives the coarse grained density. The averaging procedure is equivalent to
replacing the summation over $n$ in Eq.~(\ref{ncg1}) by integration, and we obtain the following expression for the coarse grained density:
\begin{eqnarray}
&&n_{cg}(\tilde y)=n_{2D}\left[\sqrt{\frac{2}{\pi}}\frac{1}{\tilde R^2}\int_{-\tilde R}^{\tilde R}dw(\tilde R^2-w^2)\exp\{-2(y+w)^2\}\right]   \nonumber  \\
&&=n_{2D}\sqrt{\frac{2}{\pi}}\frac{1}{\tilde R^2}\Big[\frac{\tilde R^2-\tilde y^2-1/4}{2}\sqrt{\frac{\pi}{2}}\,\left\{{\rm erf}[\sqrt{2}(\tilde y+\tilde R)]-{\rm erf}[\sqrt{2}
(\tilde y-\tilde R)]\right\} \nonumber \\
&&+\frac{\tilde R-\tilde y}{4}\exp\{-2(\tilde y +\tilde R)^2\} +\frac{\tilde R+\tilde y}{4}\exp\{-2(\tilde y -\tilde R)^2\}\Big],        \label{ncg2}
\end{eqnarray}
where $n_{2D}=3n_{1D}/4\tilde R\tilde l$ is a characteristic 2D density, and $n_{1D}=N/L\tilde l$ is the one-dimensional density in the narrow channel. The coarse grained density
versus $\tilde y$ for $\tilde R=8.9$ is displayed in Fig.~\ref{fig0}. For $(\tilde R-|\tilde y|)\gg 1$, Eq.~(\ref{ncg2}) immediately gives the expected Thomas-Fermi density
profile:
\begin{equation}        \label{ncgTF}
n_{cg}(\tilde y)=n_{2D}\left(1-\frac{\tilde y^2}{\tilde R^2}\right),
\end{equation}
and for $\tilde y=\tilde R$ we have $n_{cg}=1/\sqrt{2\pi\tilde R^2}$. As well as in the symmetric case, the inverted-parabola formula already works well not very far from the
Thomas-Fermi boarder. In Fig.~\ref{fig0} one sees that for $\tilde R=8.9$ this is the case for $|\tilde y|\leq 8$. If $|\tilde y|>\tilde R$ and $(|\tilde y|-\tilde R)\gg 1$, then
the density decays exponentially:
\begin{equation}        \label{ncgexp}
n_{cg}(\tilde y)=n_{2D}\sqrt{\frac{1}{2\pi\tilde R^2}}\,\frac{1}{4(|\tilde y|-\tilde R)^2}\exp\{-2(|\tilde y|-\tilde R)^2\}.
\end{equation}

The melting of the vortex lattice occurs  when  the number of vortices $N_v$ becomes of the order of the number of  atoms. The number of vortex rows at large $\tilde\beta$
increases as $\tilde\beta^{1/3}$, and the spacing between the vortices is $\sim \tilde l$. Thus, the number of vortices is $\sim \tilde\beta^{1/3}L$, and the melting transition
occurs at the one-dimensional density $n_{1D}=(N/L\tilde l)\sim \tilde\beta^{1/3}/\tilde l$.

\section{Phase diagram for a  condensate in the narrow channel}

In this Section we calculate the phase diagram for a rapidly rotating Bose-Einstein condensate in the narrow channel geometry. The phase diagram is obtained by numerical
minimization of the energy functional
\begin{equation}  \label{Efunctional}
 E/\hbar\omega_0=\sum_{k}{k^2}|a_{k}|^2+\tilde\beta\sum_{k,k',q}
{a}^{*}_{k+q}{a}^{*}_{k'-q}{a}_{k'}{a}_{k}\exp\left[-\frac{1}{4}\{(k-k'+q)^2+q^2\}\right],
\end{equation}
where we impose periodic boundary conditions along the $y$-axis and omit the index $n$ for momenta $k_n$. The energy is measured in units of $\hbar\omega_0$ and depends only on a
single dimensionless parameter $\tilde\beta$. The functional (\ref{Efunctional}) is obtained by substituting the condensate wavefunction
\begin{equation}
\psi(x,y)=\sum_{k}a_k\Psi_k(\tilde x,\tilde y),
\label{Psi_var}
\end{equation}
with $\Psi_k(\tilde x,\tilde y)$ given by Eq.~(\ref{setch}), into the Hamiltonian for interacting bosons in the Landau gauge and integrating over $d\tilde x$ and $d\tilde y\,$
\cite{gora}.

The coefficients $a_k$ were calculated by minimizing $E(\ref{Efunctional})$ using a simulated annealing algorithm \cite{annealing}.
In general, there is an infinite number of coefficients $a_k$ in the variational wavefunction (\ref{Psi_var}), but the ones corresponding to large momenta are strongly suppressed
due to the presence of the ``kinetic energy'' term in the energy functional. The normalization condition for the condensate wavefunction leads to the constraint $\sum_{k}{|a_k
^2}=1$.

\begin{figure}[htb]
\centering
\includegraphics[width=6.0in]{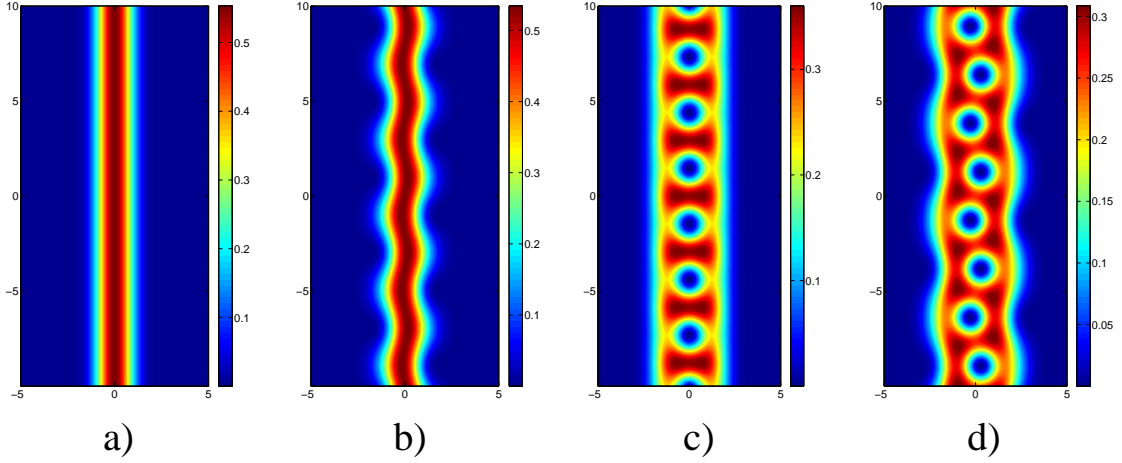}
\caption{Condensate wave-function $|\psi(x,y)|^2$ for different values of the interaction strength:
 a) $\beta=0$, b) $\beta=5.2$, c) $\beta=10$, d) $\beta=19.2.$ Coordinates $x$ (horizontal line) and $y$ (vertical line) are given in units of $\tilde l$.}
\label{phase_diag_1}
\end{figure}

\begin{figure}[htb]
\centering
\includegraphics[width=5.0in]{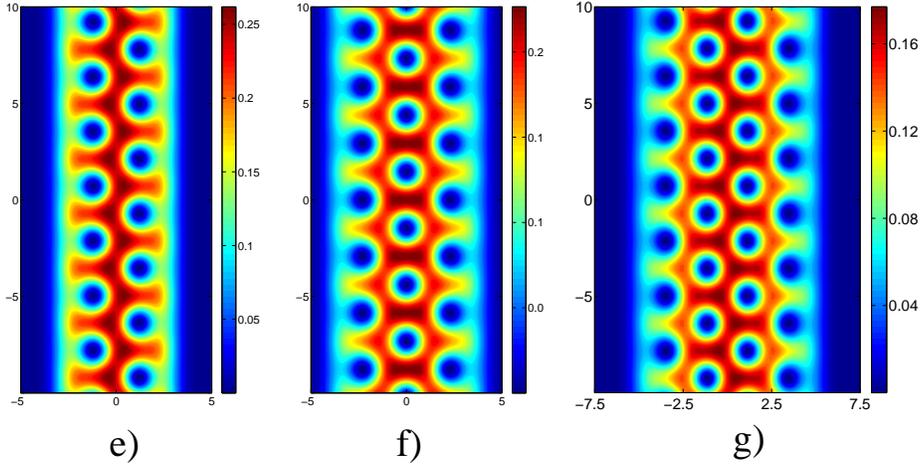}
\caption{The same as in Fig.~\ref{phase_diag_1} for: e) $\beta=30$, f) $\beta=50$, g) $\beta=100$.}
\label{phase_diag_2}
\end{figure}

At $\tilde\beta=0$ the energy is minimized by setting all coefficients $a_k$ with $k\ne0$ equal to zero and $a_0=1$.
This corresponds to the condensate density profile shown on Fig.~\ref{phase_diag_1}a, which is a
 Gaussian with the half-width $\tilde l$ in the $x$ direction and is uniform along
the $y$ axis. This state remains the ground state for §$\tilde\beta< 4.9$, and for $\tilde\beta=4.9$
 it transforms via a second order quantum phase transition into the state
displayed in Fig.~\ref{phase_diag_1}b. In this state two extra components, $k_0$ and $-k_0$, develop
 and the ordering wavevector $k_0$ has the value $k_0\simeq 2.25$ for
$\tilde\beta\simeq 5.2$. The three
main components of this state are accompanied by nonzero, but much smaller components with
 higher  $k$ which are multiples of $k_0$. The critical value of $\tilde \beta$ for this
phase transition can be obtained analytically by minimizing the energy of the three-component
 wavefunction (\ref{Psi_var}), with $k=0$ and $k=\pm k_0$ \cite{gora}. In this case
the emerging state is seen as two rows of vortices \cite{gora}, although including small components
 with higher $k$ it becomes a sort of corrugated state and can also be
identified as a density wave.

At $\tilde\beta\approx 5.4$, there is a first order phase transition from the density-wave state b) into
 the state with one vortex row (Fig.~\ref{phase_diag_1}c). In this state the
central component vanishes ($a_0=0$) and the wavefunction is characterized by two main non-zero
 components $a_{\pm k_0}\approx 1/\sqrt{2}$. At $\tilde\beta\sim 10$ the ordering
wavevector is equal to $k_0\simeq 1.51$. The energy of the purely two-component state is larger by
 a  small amount than the energy of the state c), which is especially visible near
the phase transitions.

Close to $\tilde\beta=18.9$, the state c) transforms into the state shown in Fig.~\ref{phase_diag_1}d and representing a density wave of
vortices. This looks like the first order transition (see Ref.~\cite{sanchez}). However, the state d) becomes the ground state only for $\tilde\beta>18.88$, whereas the state c) is
the ground state for $\tilde\beta<18.85$. In the narrow interval $18.85<\tilde\beta<18.88$ our
 calculations yield a dynamically unstable corrugated state d), which is signaled by
a negative sign of the compressibility. This is likely to mean that in this narrow range of $\tilde\beta$
 the system undergoes the phase separation.

\begin{figure}[htb]
\centering
\includegraphics[width=5.0in]{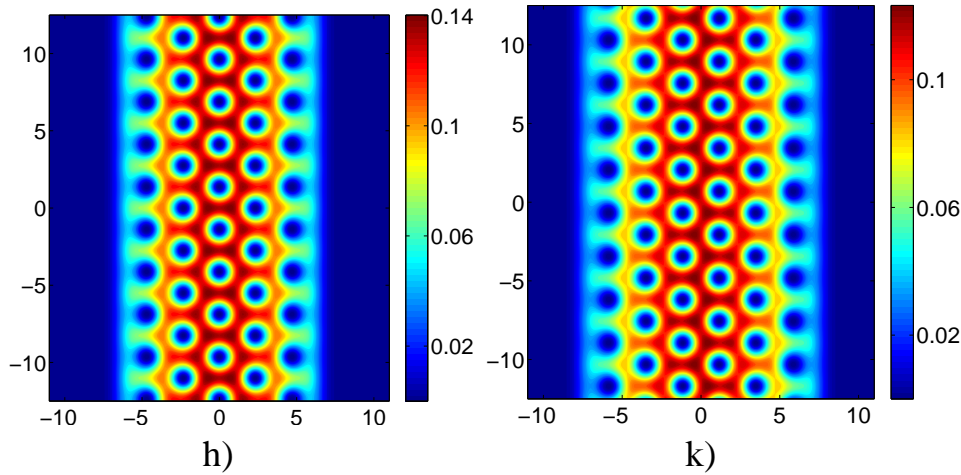}
\caption{The same as in Fig.~\ref{phase_diag_1} for: h) $\beta=200$, k) $\beta=300$.}
\label{phase_diag_3}
\end{figure}

\begin{figure}[htb]
\centering
\includegraphics[width=3.0in]{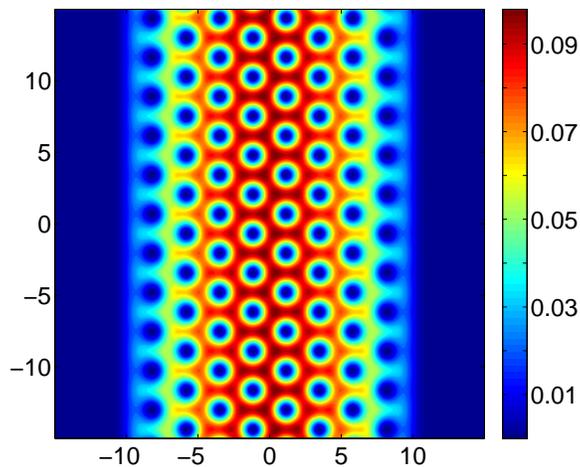}
\caption{The same as in Fig.~\ref{phase_diag_1} for: $\beta=600$ (see text).}
\label{phase_diag_4}
\end{figure}

The state d) has five main components, including the central component at $k=0$ and has the
 ordering  wavevector which is equal to $k_0\simeq 1.74$ at $\tilde\beta\simeq
19.2$. This state turns into the state with two vortex rows (Fig.~\ref{phase_diag_2}e) via a
 first-order  phase transition at $\tilde\beta\simeq 19.8$.

\begin{figure}[htb]
\centering
\includegraphics[width=6in]{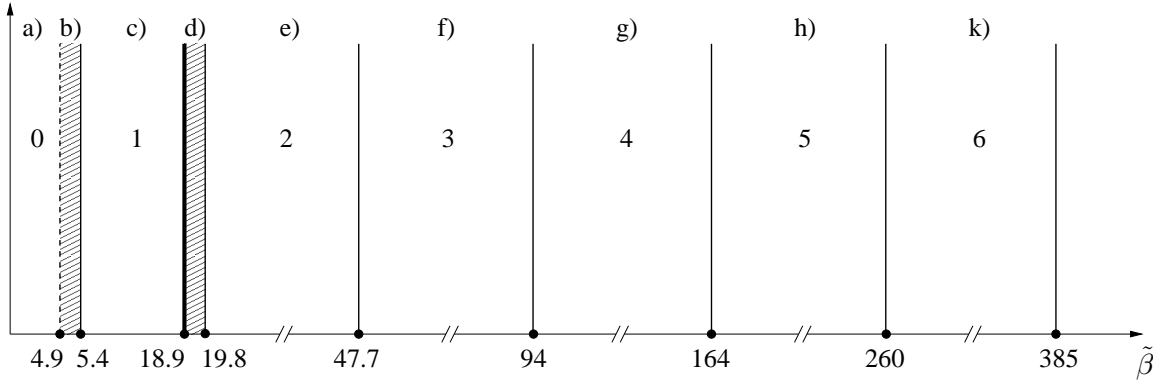}
\caption{Zero temperature phase diagram for a rapidly rotating condensate in the narrow channel. Solid vertical lines indicate the points of first order transitions, and the dashed
line the point of the second order transition. The bold solid line shows the transition between the states c) and d) (see text). The numbers from 0 to 6 stand for the number of
vortex rows in a given range of $\tilde\beta$, and the filled areas correspond to corrugated/density-wave states. The letters from a) to k) indicate the figure in which a given
vortex state is shown.}
\label{phase_diag_line}
\end{figure}

For larger $\tilde\beta$, there are phase transitions at $\tilde\beta\simeq 47.7$ to the state with three vortex rows (Fig.~\ref{phase_diag_2}f), and
at $\tilde\beta\simeq 94$ to the state with four vortex rows (Fig.~\ref{phase_diag_2}g). These transitions seem to be of the first order.
The physical explanation for the possible absence of intermediate corrugated/density-wave states near the first-order phase
transitions into the states with a larger number of vortices could be that starting from two vortex rows the system becomes
rigid to corrugations in the transverse direction.
So, increasing $\tilde\beta$ we observe an increase in the number of vortex rows through first order
transitions. For $\tilde\beta\simeq 164$ there is a transition to the state with five vortex rows, for $\tilde\beta\simeq 260$ to the state with six vortex rows, and for
$\tilde\beta\simeq 385$ to the state with seven rows, etc. (see Fig~\ref{phase_diag_3} and Fig.~\ref{phase_diag_line}). Already for the state with $8$ vortex rows, which emerges
at $\tilde\beta\simeq 550$, the composition of the rows looks like a triangular lattice (see Fig.~\ref{phase_diag_4}). For a large number $j$ of the vortex rows, the Thomas-Fermi
size of the cloud in the $x$ direction, $2\tilde R$, satisfies the asymptotic relation (\ref{tildeR}) and is proportional to $\tilde\beta^{1/3}$. It is equal to the distance
between the rows multiplied by $(j+1)$. Thus, the value of $\tilde\beta$ corresponding to the transition from $(j-1)$ to $j$ vortex rows obeys the relation
$\tilde\beta_j=\tilde\beta_{j-1}[(j+1)/j]^3$. It works with a high accuracy, which is better than $0.5\%$ for $j\geq 8$.

\begin{figure}[htb]
\centering
\includegraphics[width=5in]{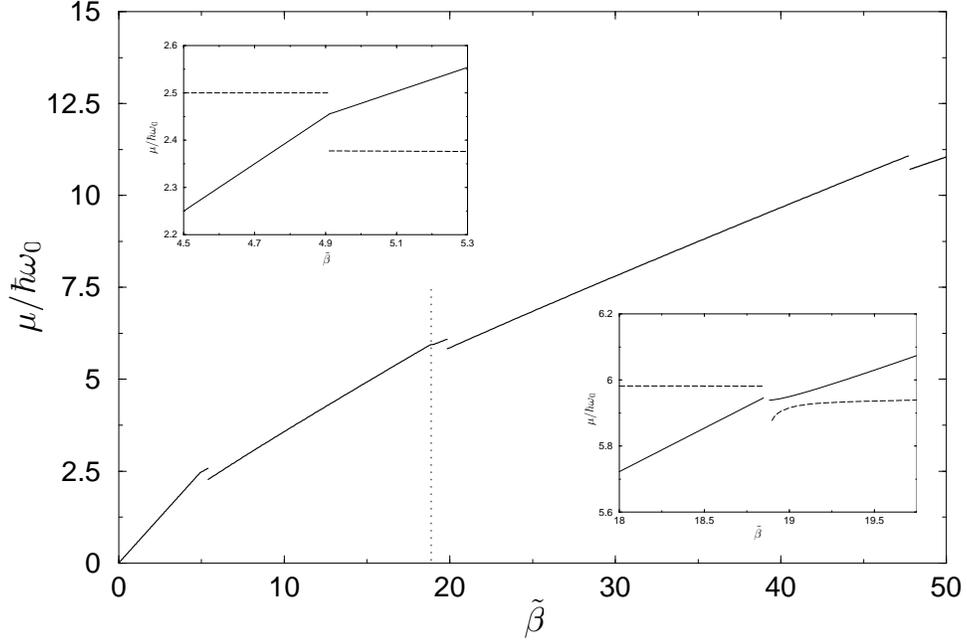}
\caption{Chemical potential in units of $\hbar\omega_0$ as a function of $\tilde\beta$. The dotted line indicates the transition between the states c) and d) (see text).
The insets show the dependence $\tilde\mu(\tilde\beta)$ in the vicinity of the quantum transitions at $\tilde\beta=4.9$ (upper inset) and at $\tilde\beta\simeq 18.9$ (lower inset).
The dashed lines in the insets indicate the derivative $\partial\tilde\mu/\partial\tilde\beta$ in arbitrary units.}
\label{chem_pot}
\end{figure}

The narrow channel geometry for rotating Bose gases was first considered in Ref.~\cite{gora},
where the roton-maxon structure of the excitations of the BEC without vortices has been found, and the
phase diagram was presented with an emphasis on the first two transitions which can be calculated
analytically. A numerical study of the related problem was done in Ref.~\cite{sanchez}, where
 corrugated states were discussed. However the analysis of quantum transitions in
  Ref.~\cite{sanchez} stops at the appearance of the state with two vortex rows, although the states
   with up to 4 rows of vortices have also
been observed. Here, we present a complete phase diagram and identify the nature of quantum
 phase  transitions. The chemical potential as a function of $\tilde\beta$ for
$\tilde\beta<50$ is shown in
Fig.~\ref{chem_pot}, indicating three first order transitions, one second order transition, and the above described peculiar transition between the states c) and d).

The extension of the condensate wavefunction in the $x$-direction increases with increasing the interaction
strength, and the two-dimensional density decreases. This decreases the average filling
factor defined as $\nu=N/N_v $. The Gross-Pitaevskii equation gives a good description in the limit of large filling factors and we
expect our picture to break down at very large $\tilde\beta$. Eventually, when the number of particles becomes of the order of the number of vortices $N_v$, the vortex lattice
melts through the phase transition to a strongly correlated state. The limiting case of extremely large $\tilde \beta$ corresponds to the Laughlin state with $\nu=1/2$, and it was
discussed for the narrow channel with periodic boundary conditions in the $x$ direction in Ref.~\cite{haldane}.

\section{Conclusion}

In conclusion, we found an analytical solution for the vortex lattice in a rapidly rotating  BEC
in the LLL. This solution is  asymptotically  exact in the limit of a very large number  of vortices, and we discuss non-Thomas-Fermi effects in the density profile.
The results  are obtained  for two  limiting cases, circularly symmetric BEC and narrow
channel geometry. In the  latter case  we present a complete phase diagram and identify the order of quantum phase
transitions occurring under an increase in the  interaction strength and/or rotation frequency and resulting in an increase in the number of vortex rows.

\section*{Acknowledgements}

We are grateful to N.R. Cooper for helpful discussions.
S.M. and S.O. acknowledge discussions with A. Aftalion, X. Blanc and F. Nier. This work was supported by ANR (Grants ANR-07-BLAN-0238 and ANR-08-BLAN-0165), by the IFRAF
Institute, and by  the Dutch Foundation FOM. G.S. wishes to thank the Aspen Center for Physics and the Institute for Nuclear Theory of the Univeresity of Washington for their
hospitality/support during the
workshop ''Quantum Simulation/Computation with Cold Atoms and Molecules'' (Aspen, May-June, 2009) and the workshop "From Femptoscience to Nanoscience: Nuclei, Quantum Dots, and
Nanostructures" (Seattle, August, 2009), where part of the present work has been done. D.K. acknowledges support from EPSRC grant EP/D066379/1. LPTMS is a mixed
research unit No.8626 of CNRS and Universit\'e Paris Sud.

\section*{Appendix}
Let us give a detailed derivation of the projection of the Gross-Pitaevskii equation for a rapidly
 rotating Bose-condensed gas onto the lowest Landau level \cite{comment}. The gas
is confined in an asymmetric harmonic potential $V({\bf r})=m(\omega_x^2x^2+\omega_y^2y^2)/2$,
 and we assume without loss of generality that $\omega_y < \omega_x$.

In the symmetric gauge the single-particle Hamiltonian is similar to that of equation (\ref{Hsingle}):
\begin{equation}
H=\frac{1}{2}(\hat {\bf p}-[{\bf \Omega}\times {\bf r}])^2+\frac{1}{ 2}(\omega_x^2-\Omega^2)x^2+ \frac{1}{2}(\omega_y^2-\Omega^2)y^2,
\label{H1}
\end{equation}
where we put $\hbar=m=1$. Drawing an analogy with a charged particle in a uniform magnetic field $B$, the rotation frequency $\Omega$ is identified with half the cyclotron
frequency $\omega_c$, and putting the particle charge and light velocity equal to unity we have $\Omega=\omega_c=B/2$. In complex coordinates the Hamiltonian (\ref{H1})rewrites as
 \begin{equation}
\label{ham} H(z,\bar z)=-2\partial\bar \partial+\omega_c(\bar z\bar \partial-z\partial)+
 \frac{1}{2}\omega_t^2z\bar z+ \frac{1}{8}(\omega_x^2-\omega_y^2)(z^2+\bar z^2),
 \end{equation}
where $\omega_t^2=\omega_x^2+\omega_y^2$ and can be rewritten as $\omega_t^2=\omega_c^2+(\tilde\omega_x^2+\tilde\omega_y^2)/2$, with
 $\tilde\omega_x^2=\omega_x^2-\omega_c^2$ and
$\tilde\omega_y^2=\omega_y^2-\omega_c^2$.

Introducing the  frequencies $\omega_t^{\pm}=\sqrt{\omega_c^2+(\frac{\tilde\omega_x\pm \tilde\omega_y}{2})^2}$ and the
dimensionless parameter
 $\rho^2=\sqrt{\frac{(\omega_t^{-}+\omega_c)(\omega_t^{+}+\omega_c)}{(\omega_t^{-}-\omega_c)(\omega_t^{+}-\omega_c) }}$,
the unnormalized ground state wavefunction is
\begin{equation}\label{a}
\langle z,\bar z|\Psi_0\rangle=e^{-\frac{1}{2}\omega_t^{+}z\bar z}e^{-\frac{1}{2}(az^2+b\bar z^2)},
\end{equation}
where $ a=\frac{1}{ 2}\rho^2(\omega_t^{-}-\omega_c)$, $ b=\frac{1}{2}(\omega_t^{-}+\omega_c)/{ \rho^2}$, and $H|\Psi_0\rangle=\omega_t^{+}|\Psi_0\rangle$.

The lowest Landau level in the asymmetric well is obtained by redefining $|\Psi\rangle=|\Psi_o\tilde{\Psi}\rangle$ and requiring the Hamiltonian $\tilde{H}$ which acts on $
\tilde{\Psi}\rangle$ to depend only on a single variable representing a linear combination of $z$ and $\bar z$:
\begin{equation}
u=z-\frac{1}{ \rho^2}\bar z.
\end{equation}
The eigenvalue equation acting on $|\tilde{\Psi}\rangle$ then reads:
\begin{equation}\label{nice}
(E-\omega_t^{+})\tilde\Psi=
\frac{2}{\rho^2}\tilde{\Psi}''+(\omega_t^{+}-\omega_t^{-})u\tilde{\Psi}'.
\end{equation}

Let us define a dimensionless variable
$u'=i\frac{\rho}{ 2}
\sqrt{\omega_t^{+}-\omega_t^{-}}\;u $
so that the eigenvalue equation (\ref{nice})  becomes
\begin{equation}(E-\omega_t^{+})\tilde\Psi=
\frac{\omega_t^{+}-\omega_t^{-}}{2}(-\tilde{\Psi}''+2u'\tilde{\Psi}')\end{equation}
This is a Hermite equation ($\omega_t^{+}>\omega_t^{-}$) with eigenfunctions $\tilde\Psi_n(u')=H_n(u')$ such that
\[\langle z,\bar z|\Psi_n\rangle=N_nH_n(u')\Psi_0(z,\bar z)\]
with eigenvalues
\[E_n=n(\omega_t^{+}-\omega_t^{-})+\omega_t^{+}.\]
Introducing the quantity $\mu$:
\[\cosh\mu=\frac{\omega_t^{-}}{\omega_c}
\sqrt{\frac{(\omega_t^{+}-\omega_c)
(\omega_t^{+}+\omega_c)}{(\omega_t^{+}-\omega_t^{-})(\omega_t^{+}+\omega_t^{-}) }}\]
\[\sinh\mu=\frac{\omega_t^{+}}{\omega_c}
\sqrt{\frac{(\omega_t^{-}-\omega_c)(\omega_t^{-}+\omega_c)}
{(\omega_t^{+}-\omega_t^{-})(\omega_t^{+}+\omega_t^{-}) }}\]
so that
\[\langle\Psi_o|z,\bar z\rangle\langle z,\bar z|\Psi_o\rangle= e^{\sinh 2\mu(-u'\bar u'
+\frac{1}{ 2}\tanh\mu(u'^2+\bar u'^2))},\]
and using the relation
\[\int \frac{du'd\bar u'}{ \pi}e^{-u'\bar u'
+\frac{1}{ 2}\tanh\mu(u'^2+\bar u'^2)}
H_l({ u'\over \sqrt{\sinh2\mu}})
H_k({ \bar u'\over \sqrt{\sinh2\mu}})=
\cosh\mu {l!\over ({\tanh\mu\over 2})^l}\delta_{k,l},\]
the normalization factor $N_n$  is found to be
\[N_n=(-i)^n{1\over\sqrt{\pi n!}}\bigg(
{\tanh\mu\over 2}\bigg)^{{n\over 2}}
\sqrt{{\omega_t^{+}\omega_t^{-}\over \omega_c}{1\over\cosh\mu}}.\]

The projector onto the LLL of an asymmetric harmonic well,
$P_{LLL}=\sum_{n\ge 0}|\Psi_n\rangle\langle\Psi_n|$, is
\begin{align}\langle z_1,\bar z_1|P_{LLL}|z_2,\bar z_2\rangle={\omega_t^{+}\omega_t^{-}\over \pi\omega_c\cosh\mu}
e^{-{1\over 2}(az_1^2+b\bar z_1^2+\omega_t^{+}z_1\bar z_1)}
e^{-{1\over 2}(a\bar z_2^2+bz_2^2+\omega_t^{+}z_2\bar z_2)}
\nonumber\\
\sum_{n\ge 0}\bigg(
{\tanh\mu\over 2}\bigg)^{n}{1\over\ n!}H_n(u_1')H_n(\bar u_2').
\end{align}
Using the relation
\[H_n(u')={2^n\over\sqrt{\pi}}\int_{-\infty}^{\infty}(u'+it)^ne^{-t^2}dt\]
one obtains
\begin{align}\langle z_1,\bar z_1|P_{LLL}|z_2,\bar z_2\rangle={\omega_t^{+}\omega_t^{-}\over \pi\omega_c}
e^{-{1\over 2}(az_1^2+b\bar z_1^2+\omega_t^{+}z_1\bar z_1)}
e^{-{1\over 2}(a\bar z_2^2+bz_2^2+\omega_t^{+}z_2\bar z_2)}\nonumber\\
e^{-\sinh^2\mu(u_1'^2+\bar u_2'^2)+\sinh2\mu u_1'\bar u_2'}.\end{align}
Any state $|\Psi_{LLL}\rangle=P_{LLL}|\Psi\rangle$ in the LLL is a linear combination of LLL eigenstates:
\[\langle z,\bar z|\Psi_{LLL}\rangle=\sum_{n=0}^{\infty}c_n\langle z,\bar z|\Psi_n\rangle
=f(u')e^{-{1\over 2}(az^2+b\bar z^2+\omega_t^{+}z\bar z)}\]
where $f(u')$ is an analytic function.

Consider now the Hamiltonian (\ref{ham}) to which we add a scalar potential $V(z,\bar z)$:
\begin{equation}
\label{hamm} H(z,\bar z)=-2\partial\bar \partial+\omega_c(\bar z\bar \partial-z\partial)
+ \frac{1}{2}\omega_t^2z\bar z+ \frac{1}{ 8}(\omega_x^2-\omega_y^2)(z^2+\bar z^2)+V(z,\bar z)
 \end{equation}
Projecting the eigenvalue equation $H|\Psi\rangle=E|\Psi\rangle$ onto the LLL amounts to
$\langle z_1,\bar z_1|P_{LLL}HP_{LLL}|\Psi\rangle=E\langle z_1,\bar z_1|P_{LLL}|\Psi\rangle$.
This gives
\begin{align}
{\omega_t^{+}\omega_t^{-}\over \pi\omega_c}\int dz_2d\bar z_2
e^{-{1\over 2}(a\bar z_2^2+bz_2^2+\omega_t^{+}z_2\bar z_2)}e^{-{1\over 2}(az_2^2+b\bar z_2^2+\omega_t^{+}z_2\bar z_2)}
e^{-\sinh^2\mu(u_1'^2+\bar u_2'^2)+\sinh2\mu u_1'\bar u_2'}
\nonumber \\H(z_2,\bar z_2)f(u_2')
=Ef(u_1').
\end{align}
Writing explicitly $H(z_2,\bar z_2)$
and changing the integration variables to $u'_2,\bar u'_2$, we have
\begin{align}\label{1}
{\sinh2\mu\over \pi}e^{-\sinh^2\mu u_1'^2}\int du_2'd\bar u_2'
e^{\sinh^2\mu u_2'^2}
e^{\sinh2\mu(u_1'\bar u_2'-u_2'\bar u_2')}\nonumber\\
\bigg(
{\omega_t^{+}-\omega_t^{-}\over 2}(-f''(u_2')+2u_2'f'(u_2'))+\omega_t^{+}f(u_2')+V(u_2',\bar u_2')f(u_2')
\bigg)
=Ef(u_1').\end{align}
Using the Bargman identity
\[{\sinh2\mu\over \pi}\int du_2'd\bar u_2'e^{\sinh2\mu(\bar u_2'u_1'-u_2'\bar u_2')}
f(u_2')=f(u_1'),\]
one finally obtains
\[(E-\omega_t^{+})f(u')={\omega_t^{+}-\omega_t^{-}\over 2}(-f''(u')+2u'f'(u'))+e^{-\sinh^2\mu u'^2}:V(u',{1\over \sinh2\mu}\partial_{u'}):e^{\sinh^2\mu u'^2}f(u'),
\]
where the notation $:V(u',{1\over \sinh2\mu}\partial_{u'}):$ means that in $V(u',\bar u')$ the variable $\bar u'$ has been replaced by the operator
${1\over \sinh2\mu}\partial_{u'}$ and the normal ordering  has been made.
In the case of the Gross-Pitaevskii equation the scalar potential is replaced by
the non-linear  term
\[g\langle\Psi_{LLL}|\Psi_{LLL}\rangle=gf(u')f(\bar u') e^{-\sinh 2\mu u'\bar u'
+\sinh^2\mu(u'^2+\bar u'^2)}\]
so that Eq.~(\ref{1}) becomes

\beqa
\label{2}
&(E-\omega_t^{+})f(u'_{1}) = \frac{\omega_t^{+}-\omega_t^{-}}{ 2}(-f''(u'_1)+2u'_1f'(u'_1))+\nonumber\\
& g\frac{\sinh2\mu}{ \pi} e^{-\sinh^2{\mu u_1'}^2}\int du_2' d\bar u_2'
e^{2\sinh^2\mu {u_2'}^2+\sinh^2{\mu \bar u_2'}^2 }
e^{2\sinh2\mu({\bar u'}_{2}u'_{1}/2-u'_{2}{\bar u'}_{2})}f^2(u'_{2})f({\bar u'}_{2}) .
\eeqa
Eq.(\ref{2}) is a general form of the non-linear Gross-Pitaevskii equation projected onto the LLL of an asymmetric harmonic trap.

Let us now concentrate on the two cases of interest, circularly symmetric geometry and narrow channel geometry. In the symmetric geometry we have
$\omega_x=\omega_y=\omega=\omega_t$,
and $\omega_t^{+}-\omega_t^{-}\to \omega_t-\omega_c$,
\[\sinh\mu\to
{\omega_t\over 2\omega_c\omega}(\tilde\omega_x-\tilde\omega_y)\to 0,\quad \cosh\mu\to 1 \]
\[-2iu'\to
\sqrt{\omega_t-\omega_c}\rho z\]
\[\rho\to
2\sqrt{{\omega\omega_c\over(\omega_t-\omega_c)(\tilde\omega_x-\tilde\omega_y)}}\to \infty\]
so that
\[u'\to i\sqrt{{\omega\omega_c\over \tilde\omega_x-\tilde\omega_y}}z\to i\infty.\]
Changing variables, $u'\to z$,
equation (\ref{2}) reduces to
\begin{equation}
\label{3}
(\omega_t-\omega_c)z_1f'(z_1)+g{\omega_t\over\pi}
\int dz_2 d\bar z_2 e^{2\omega_t(\bar z_2 z_1/2-z_2\bar z_2)}
f^2(z_2)f(\bar z_2)=(E-\omega_t)f(z_1).
\end{equation}
Finally, when $\omega\to 0$ (critical rotation), i.e. $\omega_t\to \omega_c$,  (\ref{3}) becomes
\begin{equation}\label{4}g{\omega_c\over\pi}
\int dz_2 d\bar z_2 e^{2\omega_c(\bar z_2 z_1/2-z_2\bar z_2)}
f^2(z_2)f(\bar z_2)=(E-\omega_c)f(z_1).
\end{equation}
Putting $\omega_t=\omega$, $\omega_c=\Omega$, $(E-\omega_t)=\tilde\mu$, and $f\rightarrow (\sqrt{N}/\tilde l)f$ in Eq.~(\ref{3}), we obtain Eq.~(\ref{pgp}).

In the narrow channel geometry $\tilde\omega_y\to 0$, and $\omega_t^{+}-\omega_t^{-}\to {\tilde\omega_x\tilde\omega_y/(2\omega_c)}$
\[\sinh\mu=\cosh\mu\to \sqrt{{{{\omega'_t}}({{\omega'_t}}^2-\omega_c^2)\over \omega_c\tilde\omega_x\tilde\omega_y}}\]
\[-2iu'\to\sqrt{{\tilde\omega_x\tilde\omega_y\over 2 \omega_c}}\rho u \]
\[\rho\to\sqrt{{{ {\omega'_t}}+\omega_c\over {{\omega'_t}}-\omega_c }}\]
where ${\omega'_t}=\sqrt{\omega_c^2+{\tilde\omega_x^2\over 4}}$. Changing variables, $u'\to u=z-\bar z/\rho^2$, equation (\ref{2}) reduces to
\begin{equation}\label{5}
2{{{\omega'_t}}-\omega_c\over {{\omega'_t}}+\omega_c }f''(u_1)+g{{{\omega''_t}}\over \pi}
e^{{{\omega''_t}} u_1^2/2}\int du_2d\bar u_2
e^{- {{\omega''_t}}(u_2^2+{\bar u_2}^2/2)}e^{2{{\omega''_t}}(\bar u_2u_1/2-\bar u_2u_2)}f^2(u_2)f(\bar u_2)
=(E-{{\omega'_t}})f(u_1),
\end{equation}
where ${{\omega''_t}}={{\omega'_t}}({ {{\omega'_t}}+\omega_c\over 2\omega_c})^2$. Turning to the variable $\zeta=i\sqrt{\omega''_t}u_2$ , putting $\omega_c=\Omega$, and noticing
that $\omega'_t=\tilde\Omega$ where $\tilde\Omega$ was defined in Section III, we have $2\omega''_t(\omega'_t-\omega_c)/(\omega'_t+\omega_c)=\omega_0$, with $\omega_0$ introduced
in Eq.~(\ref{1d}). Then, after rescaling the function $f$ as $f\rightarrow (\sqrt{N}/\tilde l)f$, equation (\ref{5}) transforms into Eq.~(\ref{1d}).


\begin{thebibliography}{99}
\bibitem{coop} See for review: N. R. Cooper, Advances in Physics 57, 539 (2008).
\bibitem{revfet} See for review: A. L. Fetter, Rev. Mod. Phys. {\bf 81}, 647 (2009).
\bibitem{ho} Tin-Lun Ho, Phys. Rev. Lett. {\bf 87}, 060403 (2001).
\bibitem{num1} D.A. Butts and D.S. Rokhsar, Nature {\bf 397}, 327 (1999).
\bibitem{num2} N.R. Cooper, S. Komineas, and N. Read, Phys. Rev. A {\bf 70}, 033604 (2004).
\bibitem{num3} I. Coddington, P. C. Haljan, P. Engels, V. Schweikhard, S. Tung, and E. A. Cornell, Phys. Rev. A {\bf 70}, 063607 (2004).
\bibitem{aft2} A. Aftalion, X. Blanc, J. Dalibard, Phys. Rev. A {\bf 71}, 023611 (2005).
\bibitem{rosenbusch} P. Rosenbusch, D.S. Petrov, S. Sinha, F. Chevy, V. Bretin, Y. Castin, G.V. Shlyapnikov, and J. Dalibard, Phys. Rev. Lett. {\bf 88}, 250403 (2002).
\bibitem{fet} A. L. Fetter, Phys. Rev. A {\bf 75}, 013620 (2007).
\bibitem{fet0} M. Linn, M. Niemeyer, and A.L. Fetter, Phys. Rev. A {\bf 64}, 023602 (2001).
\bibitem{oktel} M.O. Oktel, Phys. Rev. A {\bf 69}, 023618 (2004).
\bibitem{aft1} A. Aftalion, X. Blanc and F. Nier, Phys. Rev. A {\bf 73}, 011601(R) (2006).
\bibitem{aft3} A. Aftalion, X. Blanc and N. Lerner, arXiv:0804.0971.
\bibitem{gora} S. Sinha and G.V. Shlyapnikov, Phys. Rev. Lett. {\bf 94}, 150401 (2005).
\bibitem{sanchez} P. Sanchez-Lotero and J. J. Palacios, Phys. Rev. A {\bf 72}, 043613 (2005).
\bibitem{gora1} D.S. Petrov, M. Holzmann, and G.V. Shlyapnikov, Phys. Rev. Lett. {\bf 84}, 2551 (2000).
\bibitem{sw} R.A. Smith and N.K.Wilkin, Phys. Rev. A {\bf 62}, 061602(R) (2000).
\bibitem{cwg} N.R. Cooper, N.K. Wilkin, and J.M.F. Gunn, Phys. Rev. Lett. {\bf87}, 120405 (2001).
\bibitem{paredes} B. Paredes, P. Fedichev, J.I. Cirac, and P. Zoller, Phys. Rev. Lett. {\bf 87}, 010402 (2001).
\bibitem{regnault} N. Regnault, Th. Jolicoeur, Phys. Rev. B {\bf 69}, 235309 (2004).
\bibitem{mash} S. Mashkevich, S. Matveenko and S. Ouvry, Nucl. Phys. B[FS] 763 (2007) 431.
\bibitem{bargmann} v. Bargmann, Comm. Pure Appl. Math. {\bf 14}, 187 (1961); Rev. Mod. Phys. {\bf 34}, 829 (1962).
\bibitem{girvin} S. M. Girvin and T. Jach, Phys. Rev. B, {\bf 29}, 5617 (1984).
\bibitem{foot} C. Foot, private communication.
\bibitem{abramowitz} M. Abramowitz and I. A. Stegun, Handbook of Mathematical Functions (Dover, New York, 1966).
\bibitem{annealing} S. Kirkpatrick, C.D. Gelatt Jr., and M.P. Vecchi, Science {\bf 220}, 671 (1983).
\bibitem{haldane} E. H. Rezayi, F. D. M. Haldane, Phys. Rev. B {\bf 50}, 17199 (1994).
\bibitem{comment} The projected Gross-Pitaevskii equation leads to the same results as the ones obtained by using the initial equation, with the condensate wavefunction
representing a linear superposition of LLL eigenstates \cite{num1,num2,aft2,fet,oktel,aft1}.

\end{thebibliography}
\end{document}